\documentclass[aip,prb
,tightenlines
,twocolumn
]{revtex4}

\usepackage{graphicx}

\begin{document}

\title{Multibubble cavitation inception}

\author{Masato Ida}
\email[E-Mail: ]{ida.masato@jaea.go.jp}
\altaffiliation[Present address: ]{Materials and Life Science Division, J-PARC Center, Japan Atomic Energy Agency, 2-4 Shirakata-Shirane, Tokai-mura, Naka-Gun, Ibaraki 319-1195, Japan}
\affiliation{Center for Computational Science and E-systems, Japan Atomic Energy Agency, Higashi-Ueno, Taito-ku, Tokyo 110-0015, Japan}

\begin{abstract}

The inception of cavitation in multibubble cases is studied numerically and 
theoretically to show that it is different from that in single-bubble cases 
in several aspects. Using a multibubble model based on the Rayleigh-Plesset 
equation with an acoustic interaction term, we confirmed that the recently 
reported suppression of cavitation inception due to the interaction of 
non-identical bubbles can take place not only in liquid mercury but also in 
water, and we found that a relatively large bubble can significantly 
decrease the cavitation threshold pressure of a nearby small bubble. By 
examining in detail the transition region where the dynamics of the 
suppressed bubble changes drastically as the inter-bubble distance changes, 
we determined that the explosive expansion of a bubble under negative 
pressure can be interrupted and turn into collapse even though the far-field 
liquid pressure well exceeds the bubble's threshold pressure. Numerical 
results suggest that the interruption of expansion occurs when the bubble 
radius is exceeded by the instantaneous unstable equilibrium radius of the 
bubble determined using the total pressure acting on the bubble. When we 
extended the discussion to systems of larger numbers of bubbles, we found 
that a larger number of bubbles have a stronger suppression effect. The 
present findings would be useful in understanding the complex behavior of 
cavitation bubbles in practical applications where in general many 
cavitation nuclei exist and may interact with each other.

\end{abstract}

\maketitle

\section{Introduction}
\label{sec1}
Cavitation in liquids is a common phenomenon that has been observed and used 
in a wide range of fields including mechanical and chemical engineering and 
nuclear and medical applications.\cite{ref1,ref2,ref3,ref4,ref5,ref6,ref7,ref8,ref9,ref10,ref11} 
When cavitation occurs in a liquid under negative pressure, many gas or vapor 
bubbles emerge and explosively expand. The cavitation bubbles then collapse 
violently after the negative pressure is released, emitting high-speed 
liquid jets and/or shock waves. Because of its practical importance and rich 
physics, cavitation has long been studied after the pioneering work by 
Rayleigh.\cite{ref12} However, cavitation is not yet fully understood, and there 
remain many unclear aspects of even its fundamental characteristics, such as 
the inception processes, threshold pressures at which cavitation occurs, and 
lifetime of emerged bubbles.\cite{ref14,ref3,ref7,ref15,ref16}

One of the key factors in cavitation bubble dynamics is bubble-bubble 
interaction. It has long been known that, when bubbles interact with each 
other, their dynamics are sometimes significantly different from that of 
isolated bubbles. In Ref.~\onlinecite{Refadd1}, for example, Smith and 
Mesler studied 
experimentally the dynamics of a spark-generated vapor bubble interacting 
with a gas bubble in an effort to search for a method to reduce cavitation 
damage. They demonstrated that energy transfer between the bubbles alters 
their dynamics, and that positive pressure pulses emitted by the collapsing 
vapor bubble are reduced when the gas bubble stays near the vapor bubble. In 
Ref.~\onlinecite{Refadd2}, M{\o}rch studied the collapse of cavitation bubble clusters using 
an energy-balance equation and found that the collapse intensity depends on 
the volume fraction of bubbles. He also studied experimentally the 
temperature dependence of cavitation bubble dynamics and showed that the 
number of cavitation bubbles increases but their size decreases as the 
liquid temperature increases. Using a detailed theoretical model, Chahine 
and Liu\cite{Refadd3} studied the growth of a vapor bubble cluster in a superheated 
liquid, and found that the expansion rate of bubbles and the temperature 
drop at the bubble surface decrease as the number of interacting bubbles 
increases. In a numerical study of shock-wave propagation in a bubble cloud, 
Wang and Brennen\cite{ref49} showed that the expansion rate of bubbles at the 
cloud center is significantly smaller that that on the cloud surface. 
In a study 
of multibubble sonoluminescence, Mettin \textit{et al}.\cite{ref23} 
showed numerically that in a strong acoustic field, the sign of the 
secondary Bjerknes force (an interaction force acting between pulsating 
bubbles) depends in a complicated manner on the driving pressure, ambient 
radii of bubbles, and inter-bubble distance.

In this paper we examined the inception processes of cavitation in water 
in multibubble cases 
or, in other words, multi-nuclei cases. The present study was motivated by a 
recent numerical study of cavitation in liquid mercury and a technique to 
suppress it. Ida \textit{et al}.\cite{ref17,ref18} showed numerically that 
bubble-bubble 
interaction through pressure waves can have a significant impact on the 
inception of cavitation. Using a nonlinear multibubble model based on the 
Keller-Miksis equation, it was found that in certain situations, the 
explosive expansion of a bubble (i.e., cavitation inception) is completely 
suppressed by the positive pressure waves that larger neighboring bubbles 
emit when they are growing. This theoretical prediction was confirmed by an 
experiment, the results of which showed that microbubble injection into 
liquid mercury suppresses cavitation inception and significantly reduces 
cavitation damage.\cite{ref19} This finding implies that the inception 
processes of cavitation in multibubble cases can be significantly different 
from that in single-bubble cases.

We present in this paper a detailed discussion of the effect of bubble-bubble 
interaction on cavitation inception in water. The theoretical model used in 
this study is a nonlinear multibubble model 
in which Rayleigh-Plesset equations for single bubbles are 
coupled through an interaction term that represents the bubble-emitted 
pressure waves. From a number of relevant nonlinear models\cite{Refadd3,ref20,ref21,ref22,ref23,ref24,ref25,ref7,ref26} we select the classical Rayleigh-Plesset equations, which are 
for gas bubbles in an incompressible liquid, coupled through an acoustic 
interaction term not taking into account time delay effects due to the 
finite sound speed of water. The time history of the far-field liquid 
pressure is assumed to be a constant negative pressure following a 
sinusoidal decompression. When the period of the decompression process is 
set to zero, this pressure-time history is reduced to a step change that has 
been frequently considered in cavitation study. Based on the model and 
setting, we first consider the interaction of two non-identical bubbles 
under negative pressure. The numerical results suggest that the suppression 
of cavitation inception reported in Ref.~\onlinecite{ref17} can occur in water as well if 
the bubbles' ambient radii and inter-bubble distance are properly set. Then, 
we discuss several details of the results. Particular attention is focused 
on the effective cavitation pressure (a dynamic Blake threshold) of the 
suppressed bubble and the transition region in parameter space where the 
bubble's behavior drastically changes as, for example, the inter-bubble 
distance changes. We also discuss the dynamics of larger numbers of 
identical bubbles to show that they have a stronger suppression effect. In 
this discussion, we derive an exact theoretical formula that relates the 
surface velocity of interacting bubbles to their instantaneous radii. The 
theoretical formula is a direct extension of the single-bubble formula 
derived previously\cite{ref27,ref2} and has high accuracy and wide applicability. 
Using the formula we discuss how bubble-bubble interaction changes the 
expansion rate of bubbles and the negative pressure amplitude in water. The 
obtained conclusions are consistent with previous numerical and experimental 
observations. In the present investigation we found that the inception process 
of cavitation in multibubble cases can be much more complex than in 
single-bubble cases and that a variety of patterns of inception 
are possible.

The rest of this paper is organized as follows: In Sec.~\ref{sec2}, a brief 
review of cavitation bubble dynamics is presented, and 
in Sec.~\ref{sec3} the model equations used in the present study are 
introduced. Section \ref{sec4} presents numerical and theoretical results and 
discussion of these results, and Sec.~\ref{sec5} presents our concluding 
remarks, including comments about applications and future works.

\begin{figure}
\includegraphics[width=7.5cm]{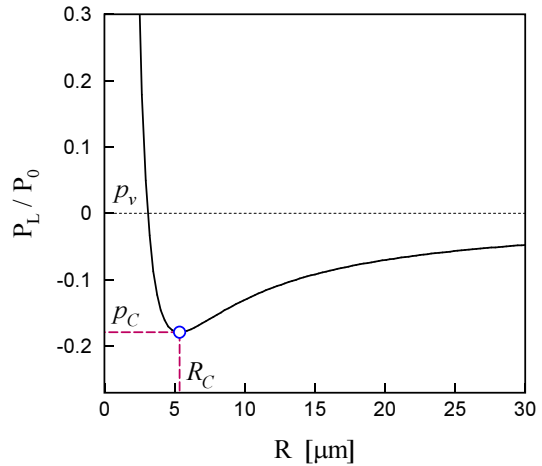}
\caption{(Color online) $p_L$--$R$ curve for $R_0 =2$ $\mu$m, $p_v =0$ Pa, 
and $\kappa =1$. The liquid pressure $p_L $ is normalized by the atmospheric 
pressure $P_0 $. $p_C $ and $R_C $ are the threshold pressure and critical 
radius, respectively.}
\label{fig1}
\end{figure}

\section{Brief review of single-bubble cavitation}
\label{sec2}
One of the well-known notions of cavitation in single-bubble cases (called 
hereafter single-bubble cavitation) in quasistatic cases is the Blake 
threshold (see, e.g., Refs.~\onlinecite{ref2,ref29,ref30}). For liquid pressures below a 
threshold value, a gas bubble (cavitation nucleus) does not have an 
equilibrium radius and thus undergoes unbounded expansion implying the 
occurrence of cavitation. The threshold liquid pressure is called the Blake 
threshold pressure and the bubble radius at the threshold is called the 
Blake critical radius. These critical values are given by the pressure 
balance equation at the bubble surface,
\begin{equation}
\label{eq1}
p_L =p_b -\frac{2\sigma }{R},
\end{equation}
where $p_L $ is the liquid pressure, $p_b $ is the internal pressure of the 
bubble given by
\begin{equation}
\label{eq2}
p_b =\left( {P_0 +\frac{2\sigma }{R_0 }-p_v } \right)\left( {\frac{R_0 }{R}} 
\right)^{3\kappa }+p_v ,
\end{equation}
$\sigma $ is the surface tension, $R$ is the bubble radius at $p_L $, $P_0 
=0.1013$ MPa is the atmospheric pressure, $R_0 $ the ambient radius of the 
bubble, $p_v $ the vapor pressure, and $\kappa $ the polytropic 
exponent of the gas inside the bubble. Equation (\ref{eq1}) describes the balance 
between $p_L $ and $p_b $ through the surface tension force $2\sigma /R$. An 
example of the $p_L$--$R$ curve given by Eq.~(\ref{eq1}) with $R_0 =2$ $\mu$m is 
shown in Fig.~\ref{fig1}. Here, we assumed that $p_v =0$ Pa for simplicity and 
$\kappa =1$, that is, the bubble interior is isothermal. This result reveals 
that for $p_L $ below a threshold value $p_C $, no bubble radius exists that 
satisfies Eq.~(\ref{eq1}). The threshold pressure, determined from $dp_L /dR=0$, is
\begin{equation}
\label{eq3}
p_C =p_v -\sqrt {\frac{32\sigma ^3}{27\left( {P_0 +\frac{2\sigma }{R_0 }-p_v 
} \right)R_0 ^3}} ,
\end{equation}
and the corresponding critical radius is
\begin{equation}
\label{eq4}
R_C =\sqrt {\frac{3}{2\sigma }\left( {P_0 +\frac{2\sigma }{R_0 }-p_v } 
\right)R_0 ^3} .
\end{equation}

When $p_L $ slowly decreases and exceeds $p_C $, the bubble begins to 
explosively expand. Then, if $p_L $ holds constant at $p_L <p_C $, the 
bubble will be in a steady growth state, where the expansion rate, $dR/dt$ 
($t$ being time), is nearly constant. The asymptotic expansion rate of such 
a bubble follows a simple formula:\cite{ref1,ref2}
\begin{equation}
\label{eq5}
\frac{dR}{dt}\approx \sqrt {\frac{2(p_v -p_L )}{3\rho }} ,
\end{equation}
where $\rho $ is the liquid density and, in this case, $p_L $ is the liquid 
pressure in the far field where the bubble-emitted pressure is negligible.

For $p_C <p_L <p_v $, Eq.~(\ref{eq1}) has two roots which correspond to 
equilibrium radii.\cite{ref27,ref28,ref29,ref30,ref2} The smaller root represents the stable 
equilibrium radius, around which the bubble can undergo damped oscillation. 
On the other hand, the larger root represents the unstable equilibrium 
radius, where a small deviation in bubble radius results in the breakdown of 
equilibrium: The bubble will shrink quickly towards the stable equilibrium 
radius if the deviation is negative but will expand without bound if the 
deviation is positive. As shown in Sec.~\ref{sec43}, the unstable equilibrium radius 
plays an important role in our study.

In dynamic cases where the liquid pressure varies rapidly and the transient 
motion of bubbles plays a role, however, the above scenario is only a rough 
description. Detailed theoretical and numerical investigations of 
single-bubble dynamics have revealed several dynamic effects on cavitation 
inception. Researchers have found, for example, that even if $\min [p_L 
(t)]>p_C $, the instantaneous bubble radius can in certain conditions be 
greater than the unstable equilibrium radius in the transient period, 
resulting in the inception of cavitation.\cite{ref27,ref28,ref31,ref32,ref29} 
This 
observation says that the effective threshold pressure and the effective 
critical radius in dynamic cases are different from those in the quasistatic 
case. Researchers also found that in a dynamic case the liquid viscosity, 
which obviously does not appear in the quasistatic formula, can alter the 
threshold pressure and the critical radius (see Ref.~\onlinecite{ref28} and references 
therein). Since in the present study we consider the effect of the pressure 
waves emitted by growing bubbles, our problem is naturally dynamic.

As mentioned above, a number of useful insights into single-bubble 
cavitation have been published. In reality, however, not only one but many 
cavitation nuclei exist and may interact with each other if they are 
sufficiently close to each other. Single-bubble study is thus not sufficient 
to fully understand cavitation dynamics in practical situations. As shown in 
Sec.~\ref{sec4}, bubble-bubble interaction can (sometimes drastically) alter 
the inception processes of cavitation in several ways.

\section{Model equations}
\label{sec3}
In the present study we assume the cavitation nuclei to be spherical gas 
microbubbles. The theoretical model used to describe their evolution is the 
coupled Rayleigh-Plesset equations, which read
\begin{equation}
\label{eq6}
R_i \ddot {R}_i +\frac{3}{2}\dot {R}_i^2 =\frac{1}{\rho }p_{s,i} 
-\sum\limits_{j=1,j\ne i}^N {\frac{1}{D_{ij} }\frac{d(R_j^2 \dot {R}_j 
)}{dt}} ,
\end{equation}
\begin{equation}
\label{eq7}
p_{s,i} =p_{b,i} -\frac{2\sigma }{R_i }-\frac{4\mu \dot {R}_i }{R_i }-p_L 
(t),
\end{equation}
\[
i=1,2,\cdots ,N,
\]
where $R_i =R_i (t)$ is the time-dependent radius of bubble $i$, $D_{ij} $ 
is the distance between the centers of bubbles $i$ and $j$, $p_L (t)$ is the 
liquid pressure in the far field, $N$ is the number of bubbles, and the 
overdots denote the time derivative $d/dt$. The surrounding liquid is 
assumed to be water of density $\rho =1000$ kg/m$^{3}$, viscosity $\mu 
=1.002\times 10^{-3}$ Pa s, and surface tension $\sigma =0.0728$ N/m. The 
compressibility of water is neglected, since we are not interested in the 
details of the collapse of bubbles. The bubble content is assumed to be an 
ideal gas, and the internal pressure of bubble $i$ ($p_{b,i} )$ is thus 
given by
\begin{equation}
\label{eq8}
p_{b,i} =\left( {P_0 +\frac{2\sigma }{R_{i0} }} \right)\left( {\frac{R_{i0} 
}{R_i }} \right)^{3\kappa _i },
\end{equation}
where $R_{i0} $ and $\kappa _i $ are the ambient radius and polytropic 
exponent, respectively, of bubble $i$. The vapor pressure and mass exchange 
across the bubble surface are neglected. We did not consider the adiabatic 
behavior of violently collapsing bubbles, since we are interested only in 
the expansion phase. Thus, $\kappa _i $ is set to unity in most cases. Since 
translation of bubbles and high-order terms depending on the translational 
velocity\cite{ref33,ref34,ref35,ref36} are neglected in this model, we set the 
inter-bubble distances $D_{ij} $ to be much greater than the ambient radii 
[e.g., as $D_{ij} \ge 10(R_{i0} +R_{j0} )$]. Using essentially the same 
model, Bremond \textit{et al}.\cite{ref26} recently showed that the coupled Rayleigh-Plesset 
equation can accurately describe the first rapid expansion and collapse, 
even when the instantaneous bubble radii reach 75{\%} of half of $D_{ij} $. 
Based on this finding, we allow the maximum bubble radii to be comparable to 
their observation.

This nonlinear system of equations describes the radial motion of $N$ 
spherical bubbles coupled through the bubble-emitted pressure waves. In this 
system, the bubbles' radial motion is driven not only by the change in $p_L 
(t)$ but also by the pressures from the neighboring bubbles described by the 
last term of Eq.~(\ref{eq6}); that is, the total driving pressure on bubble $i$ 
($p_{Ti} )$ is
\begin{equation}
\label{eq9}
p_{Ti} =p_L (t)+\rho \sum\limits_{j=1,j\ne i}^N {\frac{1}{D_{ij} 
}\frac{d(R_j^2 \dot {R}_j )}{dt}} .
\end{equation}
The last term of Eq.~(\ref{eq6}) was derived by a well-known formula for the 
pressure wave emitted by a pulsating sphere,\cite{ref37,ref23,ref17,ref2}
\begin{equation}
\label{eq10}
p_j =\frac{\rho }{r_j }\frac{d(R_j^2 \dot {R}_j )}{dt}+O\left( 
{\frac{1}{r_j^4 }} \right),
\end{equation}
where $r_j $ is the distance measured from the center of bubble $j$.

The coupling of bubbles through pressure waves, a kind of bubble-bubble 
interaction, is known to lead to a variety of phenomena that single bubbles 
can never exhibit, such as attraction and repulsion of pulsating bubbles,\cite{ref38,ref21,ref23,ref39,ref40,ref41,ref42,ref33,ref35,ref36} 
acoustic localization in bubbly liquids,\cite{ref43,ref44} 
avoided crossings of resonance frequencies,\cite{ref45} 
large phase delays,\cite{ref42} and filamentary structure formation in a strong 
sound field.\cite{ref46,ref47} In the following section we discuss the effect 
of bubble-bubble interaction on cavitation processes.

\begin{figure}
\includegraphics[width=8.2cm]{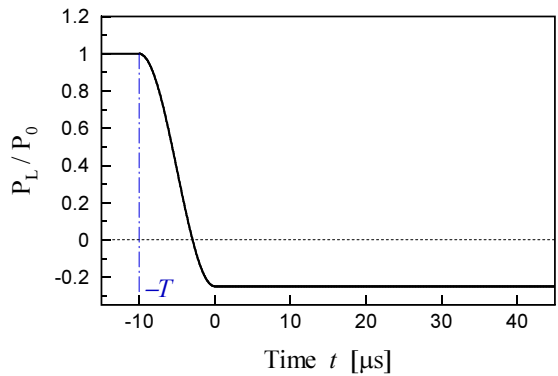}
\caption{(Color online) Pressure-time history assumed in the present study. 
The shown history is for $T=10$ $\mu$s. A similar pressure profile observed 
experimentally can be found in, e.g., Ref.~\onlinecite{ref7}.}
\label{fig2}
\end{figure}

\section{Numerical and theoretical investigations}
\label{sec4}
\subsection{Pressure-time history and initial conditions}
\label{sec41}

The time history of the far-field liquid pressure is assumed as follows:
\begin{equation}
\label{eq11}
p_L (t)=\left\{ {{\begin{array}{*{20}l}
 {P_0 } & {\mbox{for }t<-T,} \\
 {P_0 +W(P_{ng} -P_0 ) \quad} & {\mbox{for }-T\le t\le 0,} \\
 {P_{ng} } & {\mbox{otherwise},} \\
\end{array} }} \right.
\end{equation}
with
\begin{equation}
\label{eq12}
W=\frac{1-\cos \left[ {\displaystyle{ \frac{\pi }{T} }(t+T)} \right]}{2},
\end{equation}
where $P_{ng} $ is a constant negative value and $T$ is the period of the 
decompression process from $P_0 $ to $P_{ng} $. This function represents a 
constant negative pressure following a sinusoidal decompression and is 
continuous up to the first time derivative (see Fig.~\ref{fig2}). In the following 
discussions, $P_{ng} $ and $T$ are used as control parameters. When one sets 
$T\to 0$, Eq.~(\ref{eq11}) is reduced to a step change like that considered in 
Refs.~\onlinecite{ref27,ref48,ref28,ref32,ref17,ref2}.

The initial conditions assumed in the present study are
\begin{equation}
\label{eq13}
R_i (t=-T)=R_{i0} ,
\end{equation}
\begin{equation}
\label{eq14}
\dot {R}_i (t=-T)=0,
\end{equation}
that is, the bubbles are initially at equilibrium.

\begin{figure}
\includegraphics[width=7.5cm]{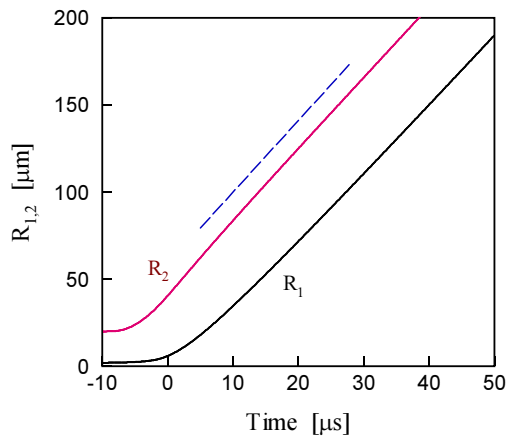}
\caption{(Color online) Radius-time curves in single-bubble cases (for 
$D_{12} \to \infty )$. The ambient radii of the bubbles are $R_{10} =2$ $\mu
$m and $R_{20} =20$ $\mu$m. The dashed line denotes the slope determined by 
Eq.~(\ref{eq5}).}
\label{fig3}
\end{figure}

\begin{figure}
\includegraphics[width=7.2cm]{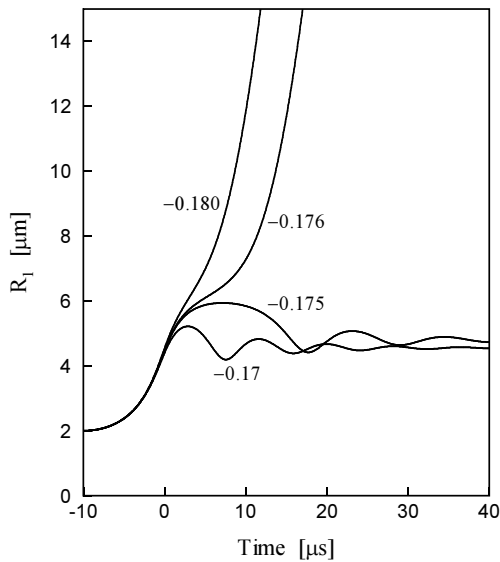}
\caption{Radius-time curves of bubble 1 for different $P_{ng} $ values 
selected around the dynamics threshold pressure. The numbers shown in the 
panel denote $P_{ng} /P_0 $.}
\label{fig4}
\end{figure}

\subsection{Competitive growth of non-identical bubbles}
\label{sec42}
Let us consider the dynamics of two non-identical bubbles (bubbles 1 and 2) 
under negative pressure. An example of a single-bubble case (i.e., for 
$D_{12} \to \infty )$ is shown in Fig.~\ref{fig3}. Here we set $R_{10} =2$ $\mu$m, 
$R_{20} =20$ $\mu$m, $\kappa _{1,2} =1$, $P_{ng} =-0.25P_0 $, and $T=10$ 
$\mu $s, and the corresponding threshold pressures are
\begin{equation}
p_{C1} =-0.179P_0 \mbox{\quad for bubble 1}
\end{equation}
and
\begin{equation}
p_{C2} =-0.007P_0 \mbox{\quad for bubble 2.}
\end{equation}
Since $P_{ng} $ well exceeds the threshold pressures, both bubbles undergo 
explosive expansion. The response of bubble 2 in the decompression process 
is in this case faster than that of bubble 1, because bubble 2 has a much 
higher threshold pressure than that of bubble 1. Hence, the explosive 
expansion of bubble 2 begins earlier than that of bubble 1. After the 
transient motion has decayed, the expansion rates of both bubbles converge 
to an almost constant value determined by Eq.~(\ref{eq5}). These observations are 
consistent with the well-known behavior of single cavitation bubbles. As 
mentioned in Sec.~\ref{sec2} the threshold pressure in dynamic cases is in general 
different from the value given by the quasistatic theory. The threshold 
pressure of bubble 1 in the present case ($T=10$ $\mu$s) is slightly higher 
than the theoretical prediction. Figure \ref{fig4} shows the radius-time curves of 
bubble 1 for different $P_{ng} $. The bubble cannot grow significantly for 
$P_{ng} \ge -0.175P_0 $, but it expands without bound for $P_{ng} \le 
-0.176P_0 $. From this observation, the dynamic threshold pressure is found 
to be about $-0.176P_0 $, which is only 1.7{\%} higher than the theoretical 
prediction. The dynamic threshold comes closer to the quasistatic prediction 
as $T$ increases.

\begin{figure}
\includegraphics[width=7.5cm]{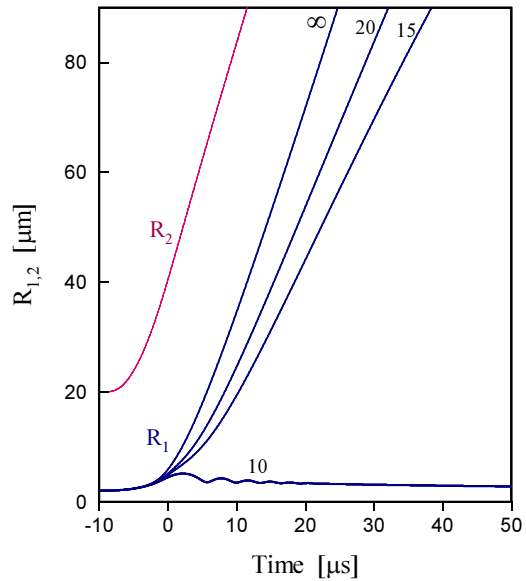}
\caption{(Color online) Radius-time curves for $P_{ng} =-0.25P_0 $ and four 
different $D_{12} $ values. The numbers denote $D_{12} /(R_{10} +R_{20} )$. 
The curves of $R_2 $ for different $D_{12} $ values are indistinguishable. 
For $D_{12} =10(R_{10} +R_{20} )$, bubble 1 cannot grow significantly 
although $p_L $ is well below the threshold pressure of the bubble.}
\label{fig5}
\end{figure}

For finite $D_{12} $, the dynamics of the bubbles can have a different 
pattern. In Fig.~\ref{fig5} we show the results for $P_{ng} =-0.25P_0 $ with four 
different values of $D_{12} $. From the figure, one finds that the expansion 
rate of bubble 1 is decreased as $D_{12} $ decreases, and the explosive 
expansion of this bubble is finally suppressed for $D_{12} =10(R_{10} 
+R_{20} )$, although the negative pressure considered here clearly exceeds 
its threshold pressure: the expansion ratio of bubble 1, $\max [R_1 
(t)]/R_{10} $, for $D_{12} =10(R_{10} +R_{20} )$ is only about 2.57. No 
considerable change occurs in the dynamics of bubble 2 in the shown period, 
because bubble 1 is too small to cause it. This result proves that the 
suppression phenomenon reported in Refs.~\onlinecite{ref17,ref18} is not inherent in liquid 
mercury but can also occur in water, whose material properties (e.g., 
density, surface tension) are greatly different from those of mercury. In 
the rest of this subsection and in the next subsection, we discuss details 
of this phenomenon.

\begin{figure}
\includegraphics[width=7.5cm]{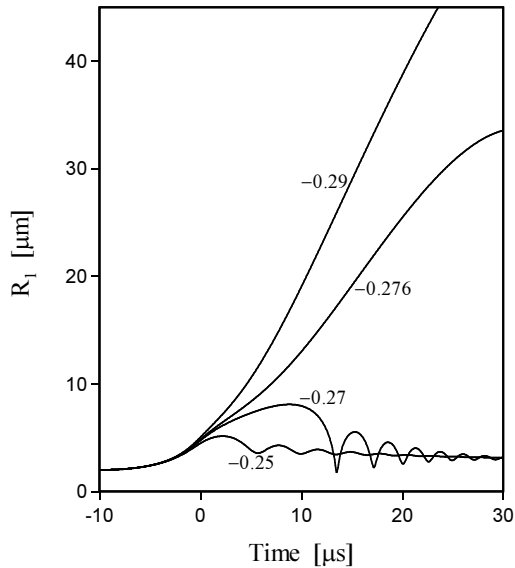}
\caption{Radius-time curves of bubble 1 for $D_{12} =10(R_{10} +R_{20} )$ and 
four different $P_{ng} $ values selected around the effective threshold 
pressure of the bubble. The numbers in the panel denote $P_{ng} /P_0 $.}
\label{fig6}
\end{figure}

The numerical result just described suggests that the \textit{effective} threshold pressure of 
bubble 1 in the case of $D_{12} =10(R_{10} +R_{20} )$ is much lower than 
that predicted by the quasistatic theory (\ref{eq3}) and a more intense negative 
pressure is thus needed to cavitate bubble 1. Figure \ref{fig6} shows the dynamics of 
bubble 1 for $D_{12} =10(R_{10} +R_{20} )$ and four different values of 
$P_{ng} $. From this, the effective threshold pressure of bubble 1 is 
deduced to be within the range of $-0.270P_0 \sim -0.276P_0 $, the absolute 
value of which is 1.5 times greater than the theoretical prediction. This 
significant change in threshold pressure is due to the positive pressure 
wave emitted by bubble 2. Bubbles expanding explosively under negative 
pressure emit positive pressure waves through their 
radial motion.\cite{ref7,ref17} 
The positive pressure waves reduce the magnitude of the negative pressure in 
the surrounding liquid, leading to the need for a more intense negative 
(far-field) pressure to cavitate. The positive pressure from bubble 2 is 
estimated by the following simple formula:\cite{ref17}
\begin{equation}
\label{eq15}
p_2 (t)=-\frac{4R_2 (t)}{3D_{12} }P_{ng} ,
\end{equation}
which is given using Eqs.~(\ref{eq10}) and (\ref{eq5}) under the assumption of $\ddot {R}_2 
\approx 0$. Since $P_{ng} $ is negative, $p_2 (t)$ is positive. The total 
pressure acting on bubble 1 is thus
\begin{equation}
\label{eq16}
P_{ng} +p_2 =\left( {1-\frac{4R_2 (t)}{3D_{12} }} \right)P_{ng} ,
\end{equation}
which is clearly higher than the far-field liquid pressure $P_{ng} $ (as 
shown in Sec.~\ref{sec44}, a larger number of bubbles cause a greater pressure 
rise). Equating this to the threshold pressure of bubble 1 ($p_{C1} )$, we 
have an approximate formula for the effective threshold pressure of bubble 
1:
\begin{equation}
\label{eq17}
P_{ng} =\frac{p_{C1} }{1-\displaystyle{ \frac{4R_2 }{3D_{12} } }}.
\end{equation}
Since the denominator of the right-hand side is smaller than unity, this 
gives a pressure lower than $p_{C1} $. This equation suggests that because 
the problem is essentially dynamic, the effective threshold pressure of a 
bubble cannot be determined uniquely but depends on the instantaneous radius 
of the neighboring bubble. Rearranging Eq.~(\ref{eq17}), one obtains a formula for 
the instantaneous radius as follows:
\begin{equation}
\label{eq18}
R_2 =\frac{3}{4}\left( {1-\frac{p_{C1} }{P_{ng} }} \right)D_{12} .
\end{equation}
For $p_{C1} =-0.179P_0 $ and $D_{12} =10(R_{10} +R_{20} )$ with $P_{ng} 
=-0.25P_0 $, this formula gives $R_2 =0.213D_{12} =46.9\mu $m. This 
criterion appears to be fulfilled in the example of Fig.~\ref{fig5}: the radius of 
bubble 2 becomes greater than this criterion before bubble 1 begins to 
expand explosively.

\begin{figure}
\includegraphics[width=8.2cm]{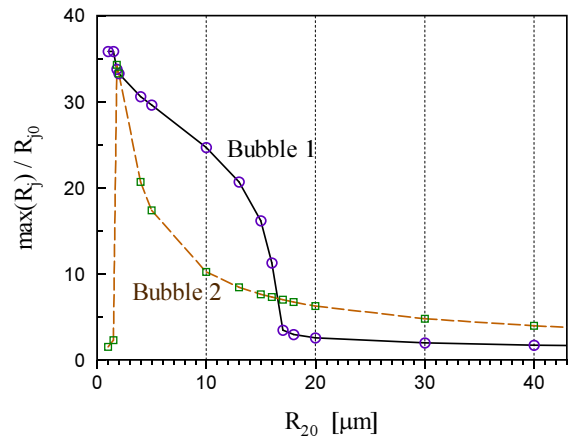}
\caption{(Color online) Expansion ratios of the bubbles as functions of 
$R_{20} $. Shown are for $R_{10} =2$ $\mu$m, $P_{ng} =-0.25P_0 $, and $t\le 
20$ $\mu$s.}
\label{fig7}
\end{figure}

\begin{figure}
\includegraphics[width=7.2cm]{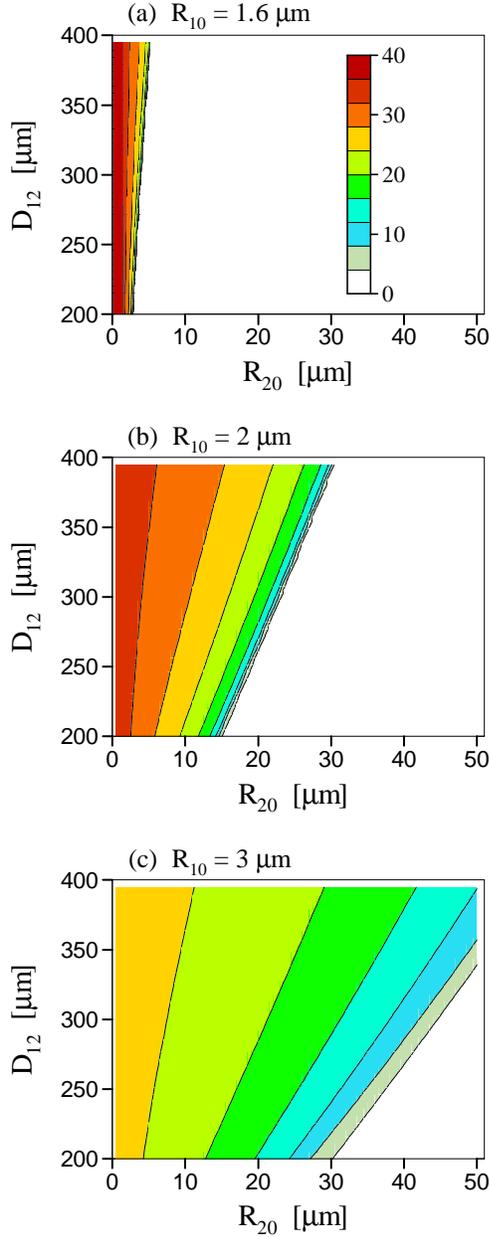}
\caption{(Color online) Phase diagrams of $\max [R_1 (t)]/R_{10} $ as 
functions of $R_{20} $ and $D_{12} $ for three different $R_{10} $ values.}
\label{fig8}
\end{figure}

The observed suppression of explosive expansion, or competitive growth of 
bubbles, occurs also for other couples of bubbles. In Fig.~\ref{fig7} we show the 
expansion ratios of bubbles, $\max [R_j (t)]/R_{j0} $, for $R_{10} =2$ $\mu
$m, $P_{ng} =-0.25P_0 $, and $t\le 20$ $\mu$s as functions of $R_{20} $. 
Here the inter-bubble distance was fixed as $D_{12} =220$ $\mu $m, 
which corresponds to $10(R_{10} +R_{20} )$ in the previous example. From 
this figure one finds that the explosive expansion of bubble 1 is completely 
suppressed when $R_{20} \ge 17$ $\mu$m. This figure also suggests that when 
$R_{20} <R_{10} $ the expansion of bubble 2 can be suppressed by bubble 1, 
and that for $R_{20} >R_{10} $ the expansion ratio of bubble 2 decreases 
monotonically as $R_{20} $ increases. The latter is due to the fact that the 
expansion rate of bubble 2 has an almost constant value determined by 
Eq.~(\ref{eq5}) although $R_{20} $ changes, and hence a larger $R_{20} $ gives a smaller 
expansion ratio. In Fig.~\ref{fig8}, we show phase diagrams of $\max [R_1 (t)]/R_{10} 
$ in a wider parameter range. Here, both $R_{20} $ and $D_{12} $ are used as 
parameters, and three different values of $R_{10} $ [(a) 1.6, (b) 2, and (c) 
3 $\mu $m] are assumed. The white regions seen in the right side of the 
panels denote cases where the explosive expansion of bubble 1 is completely 
suppressed. As can be clearly seen, the white region becomes narrower as 
$R_{10} $ increases. This is because $p_{C1} $ increases and consequently 
the pressure from bubble 2 needed for the suppression of bubble 1 becomes 
greater as $R_{10} $ increases. From the same figure, one can also see that 
larger $R_{20} $ and smaller $D_{12} $ values are more effective for 
cavitation suppression. This is consistent with the result predicted by Eq.~(\ref{eq15}) 
or (\ref{eq17}), and also with the previous numerical finding.\cite{ref17} The above 
results prove that the suppression of explosive expansion can occur if the 
bubbles' ambient radii and inter-bubble distance are appropriately set.

\begin{figure}
\includegraphics[width=7.2cm]{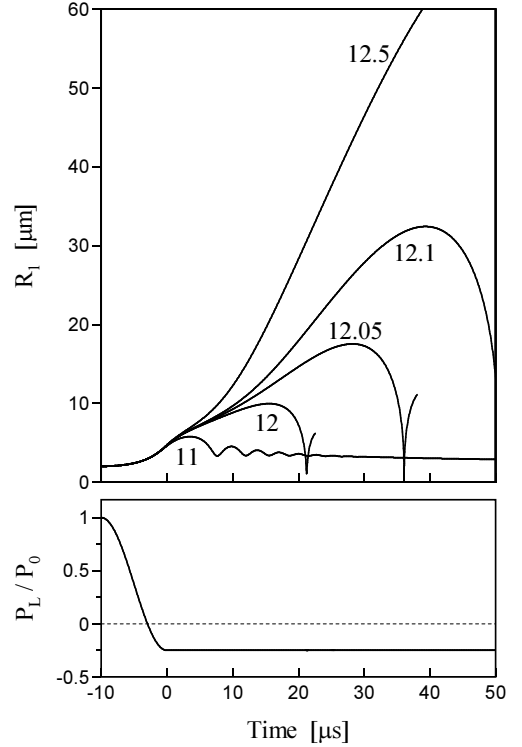}
\caption{Radius-time curves of bubble 1 for $P_{ng} =-0.25P_0 $ and five 
different $D_{12} $ values. The numbers denote $D_{12} /(R_{10} +R_{20} )$. 
For $D_{12} =12(R_{10} +R_{20} )\sim 12.1(R_{10} +R_{20} )$, the bubble 
collapses although $p_L $ (the lower panel) holds constant at a negative 
value.}
\label{fig9}
\end{figure}

\subsection{Interrupted expansion in systems of non-identical bubbles}
\label{sec43}
The results shown in Fig.~\ref{fig5} imply that a transition of bubble dynamics takes 
place in a parameter range between $D_{12} =10(R_{10} +R_{20} )$ and 
$15(R_{10} +R_{20} )$. Here we clarify what occurs in the transition region. 
In Fig.~\ref{fig9}, we show the radius-time curves of bubble 1 for $R_{10} =2$ $\mu
$m, $R_{20} =20$ $\mu$m, $P_{ng} =-0.25P_0 $, and five different $D_{12} $ 
values selected from the above-mentioned parameter range. The other 
parameters were set as in the above examples. As shown previously, 
decreasing $D_{12} $ results in the decrease of the expansion rate of bubble 
1. In the parameter range considered here, however, one more interacting 
change can be found: The expansion of bubble 1 is interrupted at a moment. 
In the results for $D_{12} =12(R_{10} +R_{20} )\sim 12.1(R_{10} +R_{20} )$, 
one finds that bubble 1 first expands considerably, but then turns into 
collapse although $p_L $ holds constant at a negative value that exceeds the 
quasistatic threshold pressure. Such a behavior is not allowed for isolated 
bubbles, which can only expand without bound. This observation suggests that 
bubble-bubble interaction sometimes causes a significant change in the 
lifetime of cavitating bubbles.

\begin{figure}
\includegraphics[width=7.5cm]{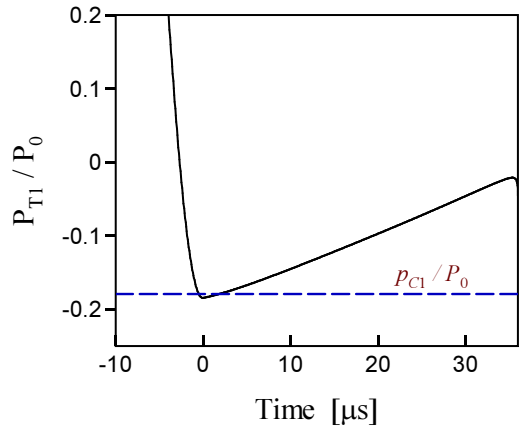}
\caption{(Color online) Total driving pressure on bubble 1 for $D_{12} 
=12.05(R_{10} +R_{20} )$. The dashed line denotes the quasistatic threshold 
pressure of bubble 1 ($p_{C1} =-0.179P_0 )$.}
\label{fig10}
\end{figure}

Let us consider the mechanism underlying this observation. As known from 
Eq.~(\ref{eq15}), the total pressure acting on bubble 1 increases as time goes on [i.e., 
as $R_2 (t)$ becomes greater]. Indeed, the total pressure at bubble 1's 
position determined numerically by
\begin{equation}
\label{eq19}
p_{T1} =p_L +\frac{\rho }{D_{12} }\frac{d(R_2^2 \dot {R}_2 )}{dt}
\end{equation}
increases for $t>0$ (see Fig.~\ref{fig10}). From this, one may presume that the 
interruption of bubble expansion occurs when the total pressure rises above 
the quasistatic threshold pressure of bubble 1. This conjecture is, however, 
incorrect or insufficient. The total pressure is clearly higher than the 
threshold pressure in most periods except for a short duration around $t=0$, 
and their crossing points are far from the time of interruption (see 
Fig.~\ref{fig10}). The notion of threshold pressure is therefore useless for the present 
purpose.

\begin{figure}
\includegraphics[width=7.0cm]{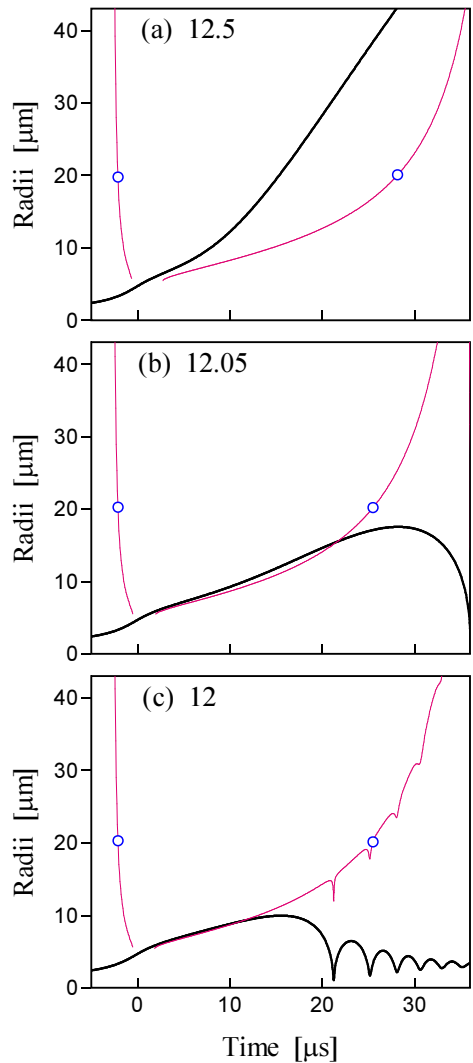}
\caption{(Color online) Radius and unstable equilibrium radius of bubble 1 
for different $D_{12} $ values as functions of time. The thick curves denote 
$R_1 $ and the thin curves with circles denote $R_{Ue1} $. The numbers 
denote $D_{12} /(R_{10} +R_{20} )$.}
\label{fig11}
\end{figure}

\begin{figure}
\includegraphics[width=7.0cm]{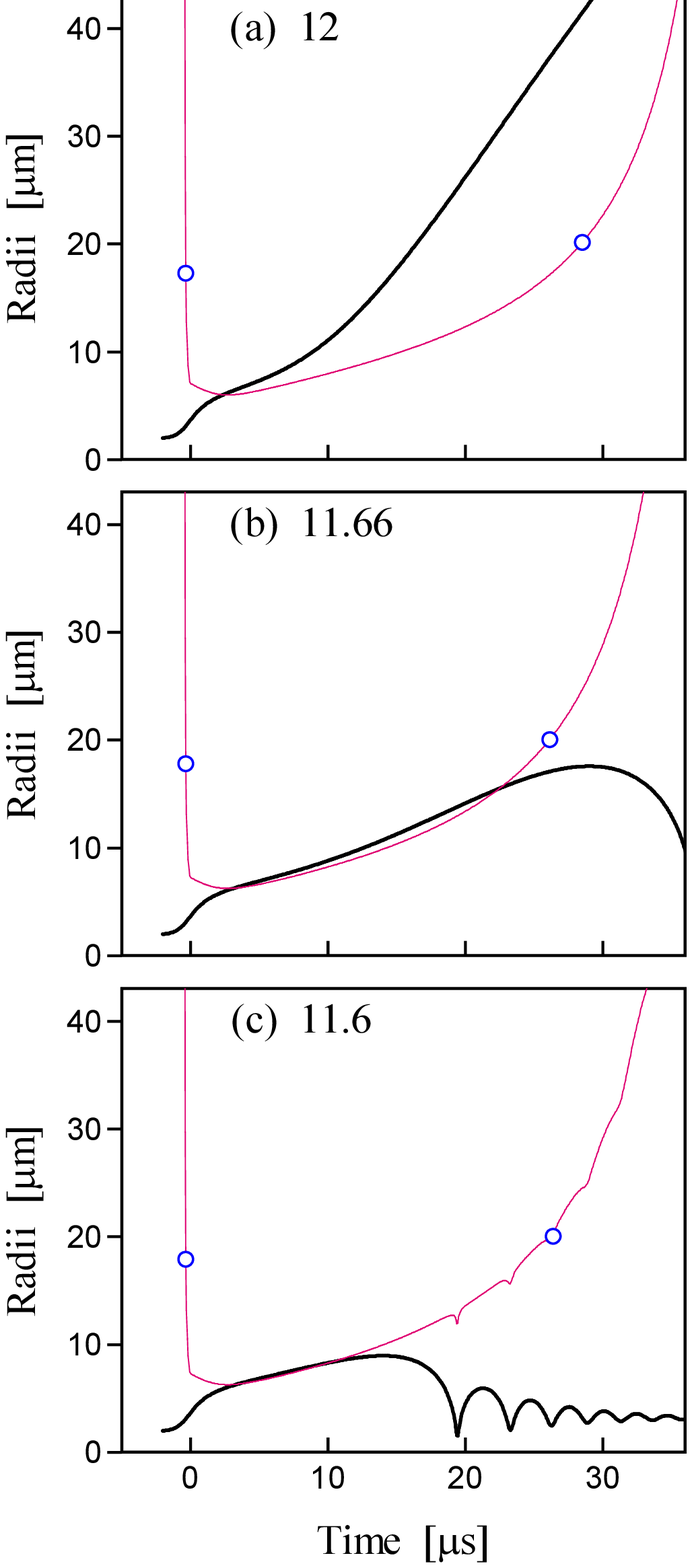}
\caption{(Color online) Same as Fig.~\ref{fig11} but for $T=2$ $\mu$s and 
smaller $D_{12} /(R_{10} +R_{20} )$.}
\label{fig12}
\end{figure}

\begin{figure}
\includegraphics[width=7.0cm]{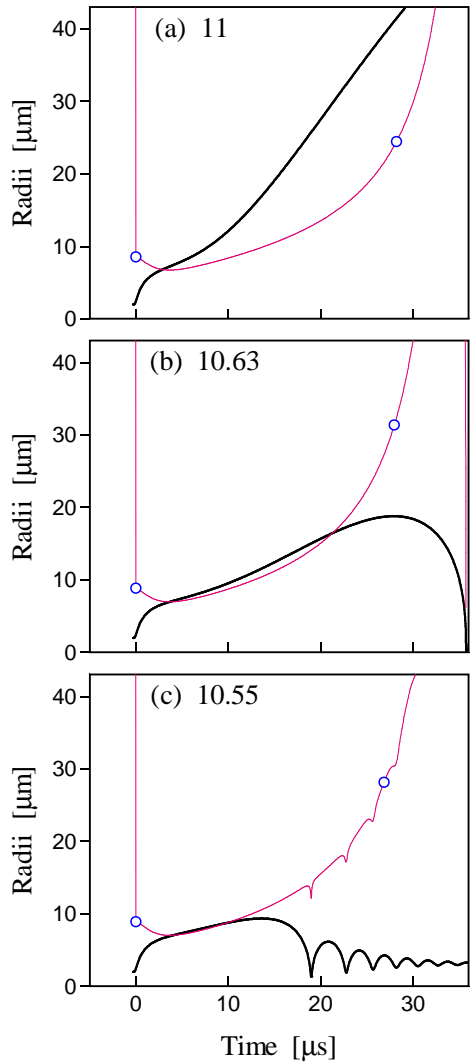}
\caption{(Color online) Same as Fig.~\ref{fig11} but for $T=0.3$ $\mu$s and 
smaller $D_{12} /(R_{10} +R_{20} )$.}
\label{fig13}
\end{figure}

We suggest using the unstable equilibrium radius to understand the 
interrupted expansion. As mentioned in Sec.~\ref{sec2}, a bubble has an unstable 
equilibrium radius under the condition of $p_C <p_L <p_v $ and begins 
unbounded expansion when the bubble's instantaneous radius becomes greater 
than the unstable equilibrium radius, even though $p_L >p_C $. We reveal 
here that the unstable equilibrium radius plays an essential role in the 
occurrence of interrupted expansion. Figure \ref{fig11} shows the radius-time curves 
of bubble 1 for three different $D_{12} $ values and the corresponding 
unstable equilibrium radius (called hereafter $R_{Ue1} )$. Here $R_{Ue1} $ 
was determined using Eq.~(\ref{eq1}) by replacing $p_L $ with $p_{T1} $ (\ref{eq19}). Since 
$p_{T1} $ is time dependent, $R_{Ue1} $ varies in time. For $D_{12} 
=12.5(R_{10} +R_{20} )$, $R_1 $ well exceeds $R_{Ue1} $ for $0$ $\mu$s 
$<t<37$ $\mu$s, and hence bubble 1 can expand rapidly as a single 
bubble does. For $D_{12} =12.05(R_{10} +R_{20} )$, $R_1 $ is slightly larger 
than $R_{Ue1} $ for $0$ $\mu$s $<t<22$ $\mu$s and bubble 1 expands 
mildly during this period. However, $R_1 $ is exceeded by $R_{Ue1} $ at 
about $t=22$ $\mu$s, and then bubble 1 stops expanding and begins 
collapsing. For $D_{12} =12(R_{10} +R_{20} )$, $R_1 $ is smaller than or 
almost equal to $R_{Ue1} $ in most periods, and hence bubble 1 cannot expand 
considerably and behaves as a stable bubble under positive absolute 
pressure. As can be seen in Fig.~\ref{fig11}(b), the crossing point of $R_1 $ with 
$R_{Ue1} $ correlates well with the timing for bubble 1 to stop 
accelerating. This observation proves that the instantaneous unstable 
equilibrium radius can be used as a probe for the interruption of bubble 
expansion.

Until now we have only considered cases of $T=10$ $\mu$s. This time length 
is greater than the characteristic time of bubble 1, $T_{b1} =0.296$ $\mu$s 
for $R_{10} =2$ $\mu$m, given by
\begin{equation}
T_{b1} =\frac{\pi }{\omega _1 }
\end{equation}
with $\omega _1 $ being the isothermal eigenfrequency of bubble 1:
\begin{equation}
\omega _1 =\sqrt {\frac{1}{\rho R_{10}^2 }\left( {3P_0 +\frac{4\sigma 
}{R_{10} }} \right)} .
\end{equation}
Here we briefly examine the effect of smaller $T$. For smaller $T$, $p_L 
(t)$ exceeds the threshold pressures of the bubbles earlier than for larger 
$T$ and the bubbles' dynamics and interaction effect would thus be altered. 
Also, as $T$ approaches $T_{b1} $, the response of bubble 1 to the 
decompression process should change. Figures \ref{fig12} and \ref{fig13} 
show results for $T=2$ $\mu
$s and $0.3$ $\mu$s ($\simeq T_{b1} )$, respectively, with $P_{ng} 
=-0.25P_0 $. In the case of $T=2$ $\mu$s, one finds that the inter-bubble 
distance of $D_{12} =12(R_{10} +R_{20} )$, which was sufficient to cause 
interrupted expansion when $T=10$ $\mu$s, is now not sufficient and a 
smaller $D_{12} $ is required. This is because the rapid expansion of bubble 
1 begins in an earlier stage than in the previous example, where the radius 
and expansion rate of bubble 2 are not so large. This result reveals that 
the dynamics of interacting bubbles depends on the time history of the 
motion. The same tendency can be found in the case of $T=0.3$ $\mu$s, where 
an even smaller $D_{12} $ is needed to interrupt the expansion of bubble 1. 
However, these results are qualitatively the same as that for $T=10$ $\mu
$s: Bubble 1 can grow considerably only when $R_1 >R_{Ue1} $.

\subsection{The case of larger numbers of bubbles}
\label{sec44}
So far we have considered double-bubble cases. In realistic cavitation, 
however, a large number of bubbles emerge at the same time and should 
interact with each other in a complicated manner. As a first step toward 
understanding the inception processes in many-bubble cases, 
in this subsection we 
discuss how a larger number of gas bubbles change the negative pressure 
amplitude in water. As shown below, a larger number of bubbles decrease the 
negative pressure amplitude more significantly and thus have a stronger 
suppression effect.

In this discussion, we categorize the bubbles considered into two groups. 
The bubbles in the first group are assumed to be very small, and their rapid 
expansion is completely suppressed by other bubbles. The bubbles in the 
second group are relatively large and can expand rapidly under an assumed 
negative pressure. This assumption allows us to neglect the impact from the 
first group on the second group, because the amplitudes of the pressure 
waves from the bubbles in the first group are negligibly small. We 
furthermore assume, as done in the local homogeneity assumption,\cite{Refadd4} that 
the bubbles in the second group are 
identical and have the same dynamics. Under these assumptions, we derive a 
theoretical formula for $\dot {R}$ of the bubbles in the second group and 
use it to examine the influence of bubble-bubble interaction on the 
expansion rate and negative pressure amplitude. Since the expansion rate 
depends on the total negative pressure on the bubble, one can see from the 
theoretical formula how the bubbles change the negative pressure value. 
The theoretical result is verified numerically for few-bubble systems using a 
test problem similar to that considered by Chahine and Liu.\cite{Refadd3}

We derive the theoretical formula for $\dot {R}$ by integrating the coupled 
Rayleigh-Plesset equation. It is known that the Rayleigh-Plesset equation 
for single bubbles can be integrated analytically if the external pressure 
is constant and the liquid viscosity is neglected.\cite{ref27,ref32,ref2} The derived 
formulas, which describe the relation between $\dot {R}(t)$ and $R(t)$, have 
provided several important insights into bubble dynamics, such as the 
asymptotic expansion rate under constant negative pressure and the minimum 
size of a bubble at collapse. Here we extend the formulas to multibubble 
systems. A theoretical formula for multibubble systems was given in 
Ref.~\onlinecite{ref7} 
under the assumptions of $\ddot {R}\approx 0$ and $R_i \gg R_{i0} $. The 
present formula is derived without these assumptions and thus has higher 
accuracy and wider applicability. We first consider a system of two 
identical bubbles and then extend to systems of larger numbers of identical 
bubbles arranged symmetrically.

When $N=2$ and $R_1 (t)=R_2 (t)$, Eq.~(\ref{eq6}) is reduced to a single 
differential equation,
\begin{equation}
\label{eq20}
R_1 \ddot {R}_1 +\frac{3}{2}\dot {R}_1^2 =\frac{1}{\rho }p_{s,1} 
-\frac{1}{D_{12} }\frac{d(R_1^2 \dot {R}_1 )}{dt},
\end{equation}
which can be rewritten as
\begin{equation}
\label{eq21}
\frac{1}{2}(2R_1 \ddot {R}_1 +3\dot {R}_1^2 )+\frac{R_1 }{2D_{12} }(2R_1 
\ddot {R}_1 +4\dot {R}_1^2 )=\frac{1}{\rho }p_{s,1} .
\end{equation}
We attempt here to transform the left-hand side of this equation to a time 
derivative. Using the identity of
\begin{equation}
\label{eq22}
\frac{1}{R_1^{m-1} \dot {R}_1 }\frac{d}{dt}(R_1^m \dot {R}_1^2 )=2R_1 \ddot 
{R}_1 +m\dot {R}_1^2 ,
\end{equation}
where $m$ is an integer, the left-hand side of Eq.~(\ref{eq21}) is rewritten as
\begin{equation}
\label{eq23}
\frac{1}{2}\frac{1}{R_1^2 \dot {R}_1 }\frac{d}{dt}\left( {R_1^3 \dot {R}_1^2 
+\frac{R_1^4 \dot {R}_1^2 }{D_{12} }} \right).
\end{equation}
From Eqs.~(\ref{eq23}) and (\ref{eq21}) we have
\begin{equation}
\label{eq24}
\frac{d}{dt}\left( {R_1^3 \dot {R}_1^2 +\frac{R_1^4 \dot {R}_1^2 }{D_{12} }} 
\right)=\frac{2R_1^2 \dot {R}_1 }{\rho }p_{s,1} .
\end{equation}
Assuming $dp_L /dt=0$ and $\mu =0$, the right-hand side of this equation can 
also be rewritten into a time derivative,
\begin{equation}
\label{eq25}
\frac{d}{dt}\left[ {-\frac{2p_L }{3\rho }R_1^3 +\frac{1}{\rho }\left( {P_0 
+\frac{2\sigma }{R_{10} }} \right)R_1^3 F-\frac{2\sigma }{\rho }R_1^2 } 
\right],
\end{equation}
where
\begin{equation}
\label{eq26}
F=\left\{ {{\begin{array}{*{20}l}
 {\frac{2}{3(1-\kappa _1 )}\left( {\frac{R_{10} }{R_1 }} \right)^{3\kappa _1 
} \quad} & {\mbox{for }\kappa _1 \ne 1,} \\
 {2\left( {\frac{R_{10} }{R_1 }} \right)^3\log R_1 } & {\mbox{for }\kappa _1 
=1.} \\
\end{array} }} \right.
\end{equation}
Substituting Eq.~(\ref{eq25}) into Eq.~(\ref{eq24}) and integrating, we finally have
\begin{eqnarray}
\label{eq27}
\left( {1+\frac{R_1 }{D_{12} }} \right)\dot {R}_1^2 &=& -\frac{2p_L }{3\rho 
}+\frac{1}{\rho }\left( {P_0 +\frac{2\sigma }{R_{10} }} \right)F \\ \nonumber
&& -\frac{2\sigma }{\rho R_1 }+\frac{\alpha }{R_1^3 }.
\end{eqnarray}
Here, $\alpha $ is the constant of integration which is determined by an 
initial condition. For $R_1 (t=0)=R_{10} $ and $\dot {R}_1 (t=0)=0$, for 
instance,
\begin{equation}
\label{eq28}
\alpha =\frac{2p_L R_{10}^3 }{3\rho }-\frac{2}{3\rho (1-\kappa )}\left( {P_0 
+\frac{2\sigma }{R_{10} }} \right)R_{10}^3 +\frac{2\sigma R_{10}^2 }{\rho }
\end{equation}
for $\kappa \ne 1$, or
\begin{equation}
\label{eq29}
\alpha =\frac{2p_L R_{10}^3 }{3\rho }-\frac{2}{\rho }\left( {P_0 
+\frac{2\sigma }{R_{10} }} \right)R_{10}^3 \log R_{10} +\frac{2\sigma 
R_{10}^2 }{\rho }
\end{equation}
for $\kappa =1$. We should note that the vapor pressure $p_v $ is neglected 
in this theory but it can easily be taken into account by only replacing 
$P_0 $, $p_L $ with $P_0 -p_v $, $p_L -p_v $, respectively.

\begin{figure}
\includegraphics[width=7.5cm]{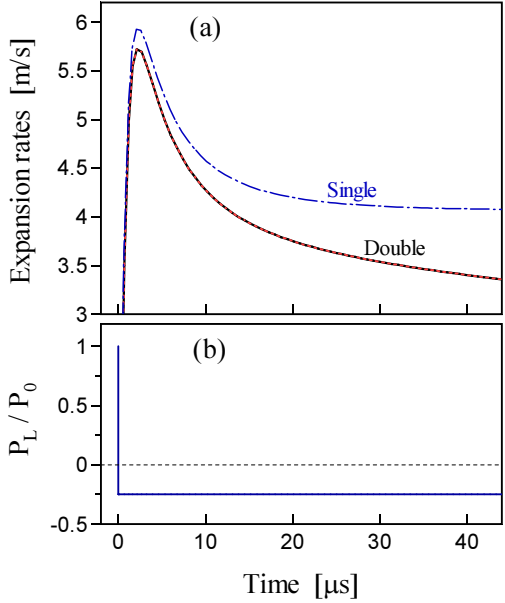}
\caption{(Color online) Expansion rates of coupled bubbles for $R_{10} =20$ 
$\mu $m, $D_{12} =20R_{10} $, and $\mu =0$ Pa s. The lower panel shows $p_L 
$ assumed in this example, normalized by $P_0 $ ($P_{ng} =-0.25P_0 $, $T=0$ 
$\mu $s). The numerical (the solid curve) and theoretical (the dots) results 
of the expansion rate are indistinguishable. The dash-dotted curve shown in 
the upper panel is the expansion rate of a single bubble with the same 
ambient radius.}
\label{fig14}
\end{figure}

\begin{figure}
\includegraphics[width=7.2cm]{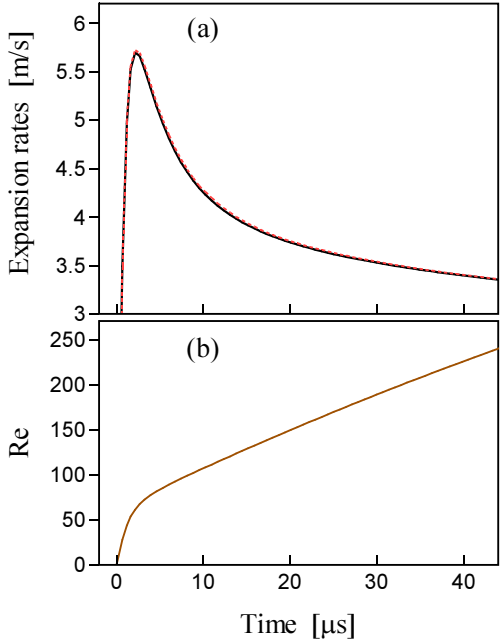}
\caption{(Color online) Same as Fig.~\ref{fig14} but for $\mu =1.002\times 10^{-3}$ 
Pa s. The lower panel shows the instantaneous Reynolds number given by 
Eq.~(\ref{eq30}).}
\label{fig15}
\end{figure}

To confirm the accuracy of the presented formula, we performed a numerical 
test. In Fig.~\ref{fig14} we show numerical and theoretical results for $R_{10} =20$ 
$\mu $m, $D_{12} =20R_{10} $, $\kappa _1 =1$, $P_{ng} =-0.25P_0 $, $\mu =0$ 
Pa s, and $T=0$ $\mu$s. The numerical results were obtained by directly 
solving the coupled Rayleigh-Plesset equation (\ref{eq20}), and the theoretical 
results were obtained using Eq.~(\ref{eq27}) with $R_1 $ given numerically and 
$\alpha $ at $t=0$ $\mu$s. These results are indistinguishable, proving the 
accuracy of Eq.~(\ref{eq27}). Figure \ref{fig14} also shows the result for $D_{12} \to \infty 
$, from which one finds that bubble-bubble interaction decreases the 
expansion rate of the bubbles. In Fig.~\ref{fig15} we show the numerical result of a 
viscous case with $\mu =1.002\times 10^{-3}$ Pa s. The result is in close 
agreement with the theory. This is because the viscosity has, in the present 
case, only a minor effect in most periods; see Fig.~\ref{fig15}(b) which reveals 
that the instantaneous Reynolds number at the bubble surface, defined here 
by
\begin{equation}
\label{eq30}
Re=\frac{\mbox{inertial term}}{\mbox{viscous term}}=\frac{\displaystyle{ \frac{3}{2}\dot {R}_i^2 }}
{\displaystyle{ \frac{4\mu \dot {R}_i }{\rho R_i} }}=\frac{3}{8}\frac{\rho R_i \dot 
{R}_i }{\mu },
\end{equation}
is large except for in the early stage.

In Eq.~(\ref{eq27}), the effect of bubble-bubble interaction is concentrated in the 
term of
\begin{equation}
\label{eq31}
\left( {1+\frac{R_1 }{D_{12} }} \right).
\end{equation}
For $D_{12} \to \infty $, this converges to unity and the formula is reduced 
to that for single bubbles.\cite{ref27,ref32,ref2} Obviously a smaller $D_{12} $ leads 
to a smaller expansion rate. This indicates that bubble-bubble interaction 
decreases the expansion rate as observed in the above numerical test. This 
conclusion is consistent with earlier numerical
observations.\cite{Refadd3,ref49,ref7,ref15,ref16} The decrease in expansion 
rate obviously indicates the reduction of the negative pressure amplitude at 
the bubbles' position. As discussed in Sec.~\ref{sec42}, this 
reduction is caused by the positive pressure waves 
emitted by the bubbles themselves;\cite{ref17} that is, bubbles expanding under 
negative pressure emit positive pressure waves that reduce the negative 
pressure amplitude in the liquid.

\begin{figure}
\includegraphics[width=7.5cm]{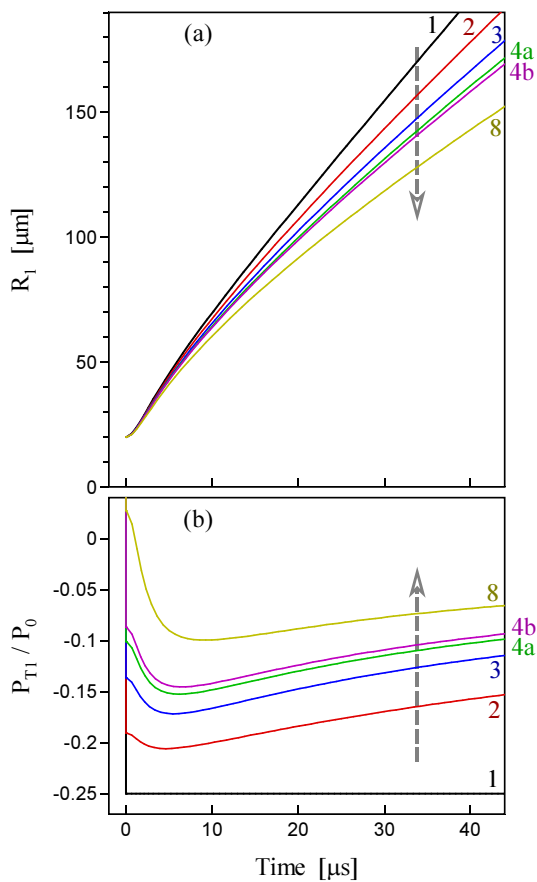}
\caption{(Color online) Bubble radii and total driving pressures on bubble 1 
for different bubble populations. The parameters are the same as those used 
in Fig.~\ref{fig14}. The numbers shown in the panels indicate the case numbers (``1'' 
and ``2'' are for the single- and double-bubble cases, respectively). As 
indicated by the arrows, the radius decreases and the total driving pressure 
increases as the bubble population increases.}
\label{fig16}
\end{figure}

Next, we briefly consider cases of larger numbers of bubbles using the 
theoretical formula. Here, following Ref.~\onlinecite{Refadd3}, we discuss 
systems of identical bubbles arranged symmetrically. This configuration 
allows us to use a model equation similar to Eq.~(\ref{eq20}). Examples 
considered are three bubbles arranged as a regular triangle (Case 3), four 
bubbles arranged as a regular tetragon (Case 4a), four bubbles arranged as a 
regular tetrahedron (Case 4b), and eight bubbles arranged as a regular 
hexahedron (Case 8). We assume that the center-to-center distances between 
nearest-neighbor bubbles (i.e., the length of the sides of the regular 
arrangements) in all cases are the same, $D_{12} $. These assumptions reduce 
Eq.~(\ref{eq6}) to a single differential equation,
\begin{equation}
\label{eq34}
R_1 \ddot {R}_1 +\frac{3}{2}\dot {R}_1^2 =\frac{1}{\rho }p_{s,1} 
-\frac{1}{E_\alpha }\frac{d(R_1^2 \dot {R}_1 )}{dt},
\end{equation}
where $E_\alpha $ is the effective inter-bubble distance given as follows:
\begin{equation}
\label{eq35}
E_\alpha =\left\{ {{\begin{array}{*{20}l}
 {E_3 =\frac{D_{12} }{2}} & {\mbox{for Case 3,}} \\
 {E_{4a} =\frac{D_{12} }{2+\frac{1}{\sqrt 2 }}} & {\mbox{for Case 4a,}} \\
 {E_{4b} =\frac{D_{12} }{3}} & {\mbox{for Case 4b,}} \\
 {E_8 =\frac{D_{12} }{3+\frac{3}{\sqrt 2 }+\frac{1}{\sqrt 3 }}} & {\mbox{for 
Case 8.}} \\
\end{array} }} \right.
\end{equation}
Since Eq.~(\ref{eq34}) has the same form as the double-bubble formula (\ref{eq20}), the 
exact formula (\ref{eq27}) can be used for the present cases by replacing $D_{12} $ 
with $E_\alpha $. From Eq.~(\ref{eq35}), one easily finds
\begin{equation}
\label{eq36}
D_{12} >E_3 >E_{4a} >E_{4b} >E_8 ,
\end{equation}
that is, a larger number of bubbles or a denser population leads to a 
stronger interaction effect and a smaller expansion rate (note that Case 4b, 
whose mean inter-bubble distance is $D_{12} $, is denser than Case 4a). This 
is confirmed by the numerical result shown in Fig.~\ref{fig16}(a) given by 
directly solving Eq.~(\ref{eq34}), which reveals the same tendency as expected. 
The present theoretical result is consistent with the numerical finding by 
Chahine and Liu.\cite{Refadd3}

The above theoretical result can also be interpreted as indicating that the 
presence of a larger number of bubbles causes a greater reduction in the 
negative pressure amplitude. In a system of a larger number of bubbles, the 
total amplitude of the bubble-emitted positive pressures should be greater 
and hence the negative pressure in the liquid, the driving force on the 
bubbles, is more reduced compared to the case of two bubbles [see 
Fig.~\ref{fig16}(b)]. This naturally results in a greater reduction of expansion rate, as 
seen above. From this argument one can conclude that a larger number of 
bubbles have a stronger suppression effect on neighboring (smaller) 
bubbles than that of single bubbles.

Although we have only considered few-bubble systems, the theoretical formula 
for $\dot {R}$ (\ref{eq27}) would be applicable to a larger cluster of bubbles. 
In recent years, there has been an effort to use a very simple model to study 
the dynamics of large bubble clusters.\cite{ref15,ref16} In those works, the degree of 
freedom of a bubble cluster is significantly reduced by a local homogeneity 
assumption\cite{Refadd4} and a model equation very similar to Eq. (\ref{eq20}) is derived. 
The present theoretical formula provides a solution of the simple model 
equation.

\section{Concluding remarks}
\label{sec5}
We have studied the inception processes of cavitation in multibubble cases, 
where multiple cavitation nuclei interact with each other, and have shown 
that bubble-bubble interaction changes the inception processes in 
a variety of ways. Performing numerical simulations of the dynamics of 
non-identical bubbles under negative pressure, we have demonstrated that the 
suppression of the explosive expansion of small bubbles by bubbles expanding 
earlier, recently reported for liquid mercury,\cite{ref17} is possible in 
water as 
well. To more deeply understand the numerical observation, we have discussed 
the effective threshold pressure of interacting bubbles. We found that even 
a bubble can significantly decrease (by about 50{\%}) the effective 
threshold pressure of a smaller neighboring bubble. This change of threshold 
value is much more significant than that caused by the dynamic effect due to 
rapid change in the far-field liquid pressure, which is only about 1.7{\%} 
in our case.

From a detailed analysis of the transition region where the dynamics of the 
suppressed bubble drastically changes as the inter-bubble distance changes, 
we have revealed that the explosive expansion of a bubble under negative 
pressure can be interrupted and turn into collapse even though the far-field 
liquid pressure remains well below the bubble's (quasistatic) threshold 
pressure. Using the notion of unstable equilibrium radius, we have found 
that the interruption of bubble expansion takes place when the instantaneous 
bubble radius is exceeded by the instantaneous unstable equilibrium radius 
determined using the total pressure acting on the bubble. 
Both the suppression and interruption of bubble 
expansion are caused by the pressure wave that a neighboring bubble emits 
when it grows.

By analytically integrating the acoustically coupled Rayleigh-Plesset 
equations under the assumption of constant far-field liquid pressure, we 
have derived an exact formula for the surface velocity of interacting 
identical bubbles. The formula shows clearly that bubble-bubble interaction 
decreases the expansion rate of bubbles and that the decrease is more 
prominent for systems of a larger number of bubbles. This theoretical result 
is consistent with earlier numerical findings.\cite{Refadd3} This 
change in the expansion rate implies that the presence of a larger number of 
bubbles reduces the negative pressure amplitude in water to a greater degree 
than in the case with two bubbles, and hence a larger number of bubbles have 
a stronger suppression effect. The presented exact formula would be applicable 
to a large bubble cluster by incorporating the local homogeneity 
assumption.\cite{Refadd4}

The present findings could be a key to understanding the complex behavior of 
cavitation bubbles in practical situations where a large number of 
cavitation nuclei exist and interact with each other. From the present 
results, one can for example speculate as follows: Even if a liquid involves 
a large number of cavitation nuclei and one imposes a strong negative 
pressure that can cavitate all of the individual nuclei, only part of them 
will grow and be observed because bubbles growing earlier suppress the 
expansion of the remaining neighboring nuclei. If this scenario is true, 
prediction of the actual population of cavitation bubbles in practical 
situations will require that, among other things, one know not only the 
initial distribution of cavitation nuclei but also their dynamic behavior in 
which bubble-bubble interaction plays a significant role. Also, the present 
findings may be relevant to the study of the \textit{superstability} of nanobubbles,\cite{ref51} where the 
interaction between nanobubbles on a hydrophobic surface and cavitation 
microbubbles is seen. In that study, Borkent \textit{et al}.~found experimentally that 
surface nanobubbles do not cavitate even for a sufficiently strong negative 
pressure, while cavitation bubbles originating from microscopic cracks are 
rapidly expanding.

More detailed analysis of multibubble cavitation dynamics by bifurcation and 
perturbation theories or other theories that have been used for 
single-bubble study\cite{ref27,ref28,ref31,ref29,ref30} would be an interesting subject to 
pursue. Also, detailed parametric study of the effective threshold pressure 
in multibubble cases could provide useful insights into the cavitation 
threshold pressure in practical applications. Statistical analysis of the 
actual population of explosively expanding bubbles in cases where 
non-identical nuclei interact would also be a meaningful area of study. 
Combining the presented pictures of cavitation inception with a scenario of 
nuclei merging, like that proposed by Marschall 
\textit{et al}.,\cite{ref14} may provide a 
more realistic picture of multibubble cavitation inception.

\acknowledgments
{
This work was partly supported by the Ministry of Education, Culture, 
Sports, Science, and Technology of Japan (MEXT) through a Grant-in-Aid for 
Young Scientists (B) (No.~20760122).
}


\begin{thebibliography}{}

\bibitem{ref1} M. S. Plesset and A. Prosperetti, ``Bubble dynamics and cavitation,'' 
Ann. Rev. Fluid Mech. \textbf{9}, 145 (1977).

\bibitem{ref2} C. E. Brennen, \textit{Cavitation and Bubble Dynamics} (Oxford University Press, New York, 1995).

\bibitem{ref3} F. Caupin and E. Herbert, ``Cavitation in water: a review,'' C. R. 
Physique \textbf{7}, 1000 (2006).

\bibitem{ref4} K. S. Suslick, ``Sonochemistry,'' Science \textbf{247}, 1439 (1990).

\bibitem{ref5} L. H. Thompson and L. K. Doraiswamy, ``Sonochemistry: Science and 
engineering,'' Ind. Eng. Chem. Res. \textbf{38}, 1215 (1999).

\bibitem{ref6} B. Riemer, J. Haines, M. Wendel, G. Bauer, M. Futakawa, S. Hasegawa, and 
H. Kogawa, ``Cavitation damage experiments for mercury spallation targets at 
the LANSCE -- WNR in 2005,'' J. Nucl. Mater. \textbf{377}, 162 (2008).

\bibitem{ref7} M. Ida, T. Naoe, and M. Futakawa, ``Direct observation and theoretical 
study of cavitation bubbles in liquid mercury,'' Phys. Rev. E \textbf{75}, 
046304 (2007).

\bibitem{ref8} T. Lu, R. Samulyak, and J. Glimm, ``Direct Numerical Simulation of 
Bubbly Flows and Application to Cavitation Mitigation,'' J. Fluids Eng. 
\textbf{129} (2007) 595.

\bibitem{ref9} E. Robert, J. Lettry, M. Farhat, P. A. Monkewitz, and F. Avellan, 
``Cavitation bubble behavior inside a liquid jet,'' Phys. Fluids 
\textbf{19}, 067106 (2007).

\bibitem{ref10} D. L. Miller, S. V. Pislaru, and J. F. Greenleaf, ``Sonoporation: 
Mechanical DNA delivery by ultrasonic cavitation,'' Somat. Cell Mol. Genet. 
\textbf{27}, 115 (2002).

\bibitem{ref11} C. C. Coussios and R. A. Roy, ``Applications of acoustics and 
cavitation to noninvasive therapy and drug delivery,'' Ann. Rev. Fluid Mech. 
\textbf{40}, 395 (2008).

\bibitem{ref12} L. Rayleigh, ``On the pressure developed in a liquid during the 
collapse of a spherical cavity,'' Philos. Mag. \textbf{34}, 94 (1917).

\bibitem{ref14} H. B. Marschall, K. A. M{\o}rch, A. P. Keller, and M. Kjeldsen, 
``Cavitation inception by almost spherical solid particles in water,'' Phys. 
Fluids \textbf{15}, 545 (2003).

\bibitem{ref15} M. Arora, C. D. Ohl, and D. Lohse, ``Effect of nuclei concentration on 
cavitation cluster dynamics,'' J. Acoust. Soc. Am. \textbf{121}, 3432 
(2007).

\bibitem{ref16} K. Yasui, Y. Iida, T. Tuziuti, T. Kozuka, and A. Towata, ``Strongly 
interacting bubbles under an ultrasonic horn,'' Phys. Rev. E \textbf{77}, 
016609 (2008).

\bibitem{Refadd1} R. H. Smith and R. B. Mesler, ``A photographic study of the effect 
of an air bubble on the growth and collapse of a vapor bubble near a 
surface,'' ASME J. Basic Eng. \textbf{94}, 933 (1972).

\bibitem{Refadd2} K. A. M{\o}rch, ``Energy considerations on the collapse of cavity 
clusters,'' Appl. Sci. Res. \textbf{38}, 313 (1982).

\bibitem{Refadd3} G. L. Chahine and H. L. Liu, ``A singular-perturbation theory of 
the growth of a bubble cluster in a superheated liquid,'' J. Fluid Mech. 
\textbf{156}, 257 (1985).

\bibitem{ref49} Y.-C. Wang and C. E. Brennen, ``Numerical computation of shock waves in 
a spherical cloud of cavitation bubbles,'' J. Fluids Eng. \textbf{121}, 872 
(1999).

\bibitem{ref23} R. Mettin, I. Akhatov, U. Parlitz, C. D. Ohl, and W. Lauterborn, 
``Bjerknes forces between small cavitation bubbles in a strong acoustic 
field,'' Phys. Rev. E \textbf{56}, 2924 (1997).

\bibitem{ref17} M. Ida, T. Naoe, and M. Futakawa, ``Suppression of cavitation inception 
by gas bubble injection: A numerical study focusing on bubble-bubble 
interaction,'' Phys. Rev. E \textbf{76}, 046309 (2007).

\bibitem{ref18} M. Ida, T. Naoe, and M. Futakawa, ``On the effect of microbubble injection on cavitation bubble dynamics in liquid mercury,'' 
Nucl. Instr. and Meth. A \textbf{600}, 367 (2009).

\bibitem{ref19} T. Naoe, M. Ida, and M. Futakawa, ``Cavitation damage reduction by 
microbubble injection,'' Nucl. Instr. and Meth. A \textbf{586}, 382 (2008).

\bibitem{ref20} A. Shima, ``The natural frequencies of two spherical bubbles 
oscillating in water,'' ASME J. Basic Eng. \textbf{93}, 426 (1971).

\bibitem{ref21} E. A. Zabolotskaya, ``Interaction of gas bubbles in a sound field,'' 
Sov. Phys. Acoust. \textbf{30}, 365 (1984).

\bibitem{ref22} S. Fujikawa and H. Takahira, ``A theoretical study on the interaction 
between two spherical bubbles and radiated pressure waves in a liquid,'' 
Acustica \textbf{61}, 188 (1986).

\bibitem{ref24} M. Ida, ``A characteristic frequency of two mutually interacting gas 
bubbles in an acoustic field,'' Phys. Lett. A \textbf{297}, 210 (2002).

\bibitem{ref25} A. Ooi and R. Manasseh, ``Coupled nonlinear oscillations of 
microbubbles,'' ANZIAM J. \textbf{46}(E), C102, (2005).

\bibitem{ref26} N. Bremond, M. Arora, S. M. Dammer, and D. Lohse, ``Interaction of 
cavitation bubbles on a wall,'' Phys. Fluids \textbf{18}, 121505 (2006).

\bibitem{ref27} J. T. S. Ma and P. K. C. Wang, ``Effect of initial air content on the 
dynamics of bubbles in liquids,'' IBM J. Res. Dev. \textbf{6}, 472 (1962).

\bibitem{ref29} A. Harkin, A. Nadim, and T. J. Kaper, ``On acoustic cavitation of 
slightly subcritical bubbles,'' Phys. Fluids \textbf{11}, 274 (1999).

\bibitem{ref30} Z. C. Feng and L. G. Leal, ``Nonlinear bubble dynamics,'' Annu. Rev. 
Fluid. Mech. \textbf{29}, 201 (1997).

\bibitem{ref28} H.-C. Chang and L.-H. Chen, ``Growth of a gas bubble in a viscous 
fluid,'' Phys. Fluids \textbf{29}, 3530 (1986).

\bibitem{ref31} A. J. Szeri and L. G. Leal, ``The onset of chaotic oscillations and 
rapid growth of a spherical bubble at subcritical conditions in an 
incompressible liquid,'' Phys. Fluids A \textbf{3}, 551 (1991).

\bibitem{ref32} C. Dugu\'{e}, D. H. Fruman, J-Y. Billard, and P. Cerrutti, ``Dynamic 
criterion for cavitation of bubbles,'' ASME J. Fluids Eng. \textbf{114}, 250 
(1992).

\bibitem{ref33} H. N. Oguz and A. Prosperetti, ``A generalization of the impulse and 
virial theorems with an application to bubble oscillations,'' J. Fluid Mech. 
\textbf{218}, 143 (1990).

\bibitem{ref34} H. Takahira, T. Akamatsu, and S. Fujikawa, ``Dynamics of a cluster of 
bubbles in a liquid: theoretical analysis,'' JSME Int. J. Ser. B 
\textbf{37B}, 297 (1994).

\bibitem{ref35} A. Harkin, T. J. Kaper, and A. Nadim, ``Coupled pulsation and 
translation of two gas bubbles in a liquid,'' J. Fluid Mech. \textbf{445}, 
377 (2001).

\bibitem{ref36} A. A. Doinikov, ``Translational motion of two interacting bubbles in a 
strong acoustic field,'' Phys. Rev. E \textbf{64}, 026301 (2001).

\bibitem{ref37} Z. Ye, ``A note on nonlinear radiation from a gas bubble in liquids,'' 
J. Acoust. Soc. Am. \textbf{101}, 809 (1997).

\bibitem{ref38} L. A. Crum, ``Bjerknes forces on bubbles in a stationary sound field,'' 
J. Acoust. Soc. Am. \textbf{57}, 1363 (1975).

\bibitem{ref39} A. A. Doinikov, ``Viscous effects on the interaction force between two 
small gas bubbles in a weak acoustic field,'' J. Acoust. Soc. Am. 
\textbf{111}, 1602 (2002).

\bibitem{ref40} M. Ida, ``Alternative interpretation of the sign reversal of secondary 
Bjerknes force acting between two pulsating gas bubbles,'' Phys. Rev. E 
\textbf{67}, 056617 (2003).

\bibitem{ref41} M. Ida, ``Investigation of transition frequencies of two acoustically 
coupled bubbles using a direct numerical simulation technique,'' J. Phys. 
Soc. Jpn. \textbf{73}, 3026 (2004).

\bibitem{ref42} M. Ida, ``Phase properties and interaction force of acoustically 
interacting bubbles: A complementary study of the transition frequency,'' 
Phys. Fluids \textbf{17}, 097107 (2005).

\bibitem{ref43} Z. Ye and A. Alvarez, ``Acoustic localization in bubbly liquid media,'' 
Phys. Rev. Lett. \textbf{80}, 3503 (1998).

\bibitem{ref44} M. Kafesaki, R. S. Penciu, and E. N. Economou, ``Air bubbles in water: 
A strongly multiple scattering medium for acoustic waves,'' Phys. Rev. Lett. 
\textbf{84}, 6050 (2000).

\bibitem{ref45} M. Ida, ``Avoided crossings in three coupled oscillators as a model 
system of acoustic bubbles,'' Phys. Rev. E \textbf{72}, 036306 (2005).

\bibitem{ref46} U. Parlitz, C. Scheffczyk, I. Akhatov, and W. Lauterborn, ``Structure 
formation in cavitation bubble fields,'' Chaos, Solitons {\&} Fractals 
\textbf{5}, 1881 (1995).

\bibitem{ref47} I. Akhatov, U. Parlitz, and W. Lauterborn, ``Towards a theory of 
self-organization phenomena in bubble-liquid mixtures,'' Phys. Rev. E 
\textbf{54}, 4990 (1996).

\bibitem{ref48} Y. Matsumoto and A. E. Beylich, ``Influence of homogeneous condensation 
inside a small gas bubble on its pressure response,'' J. Fluids Eng. 
\textbf{107}, 281 (1985).

\bibitem{Refadd4} A. Kubota, H. Kato, and H. Yamaguchi, ``A new modelling of cavitating flows: a numerical study of unsteady cavitation on a hydrofoil section,'' J. Fluid Mech. \textbf{240}, 59 (1992).

\bibitem{ref51} B. M. Borkent, S. M. Dammer, H. Schonherr, G. J. Vancso, and D. Lohse, 
``Superstability of surface nanobubbles,'' Phys. Rev. Lett. \textbf{98}, 
204502 (2007).

\end{thebibliography}
\end{document}